\documentclass[journal]{IEEEtran}

\IEEEoverridecommandlockouts

\usepackage{tikz}
\usepackage{textcomp}
\usepackage{lipsum}

\newcommand\copyrighttext{%
  \footnotesize
  \textit{N. Pompeo, K. Torokhtii, A. Alimenti, A. Mancini, G. Celentano, E. Silva, IEEE Trans. Appl. Supercond., accepted for publication (2019)}  \vspace{.2cm}\\
  \textcopyright 2019 IEEE. Personal use of this material is permitted.
  Permission from IEEE must be obtained for all other uses, in any current or future
  media, including reprinting/republishing this material for advertising or promotional
  purposes, creating new collective works, for resale or redistribution to servers or
  lists, or reuse of any copyrighted component of this work in other works.
  }
\newcommand\copyrightnotice{%
\begin{tikzpicture}[remember picture,overlay]
\node[anchor=north,yshift=-10pt] at (current page.north) {\fbox{\parbox{\dimexpr\textwidth-\fboxsep-\fboxrule\relax}{\copyrighttext}}};
\end{tikzpicture}%
}

\ifCLASSINFOpdf

\else
  \fi

\usepackage{graphicx}
\usepackage{amsmath}
\usepackage{amssymb}
\usepackage{mathrsfs}

\newcommand{\rmi}{ \mathrm{i} }

\usepackage{color}
\usepackage{nicefrac}
\usepackage{xspace}
\usepackage{color}
\usepackage[normalem]{ulem}

\newcommand{\textoutr}[1]{}

\newcommand{\Zsub}{Z^{(sub)}_{eff}}
\newcommand{\Rsub}{R^{(sub)}_{eff}}
\newcommand{\rhos}{\tilde\rho_s}

\vspace{-1.0em}
\newcommand*{\myscalefactor}{0.48}

\begin{document}

\newcommand{\IEEEmw}{{\it IEEE Trans. Microwave Theory Tech. }}
\newcommand{\IEEEas}{{\it IEEE Trans. Appl. Supercond.}}
\newcommand{\IEEEim}{{\it IEEE Trans. Instr. Meas.}}
\newcommand{\PRB}{{\it Phys. Rev. B}}
\newcommand{\PRL}{{\it Phys. Rev. Lett.}}
\newcommand{\PR}{\it Phys. Rev.}
\newcommand{\PL}{\it Phys. Lett.}
\newcommand{\APL}{{\it Appl. Phys. Lett.}}
\newcommand{\JAP}{{\it J. Appl. Phys.}}
\newcommand{\SUST}{{\it Supercond. Sci. Technol.}}
\newcommand{\etal}{{\it et al}}

\setlength{\floatsep}{5pt plus 2pt minus 2pt}
\setlength{\textfloatsep}{5pt plus 2pt minus 2pt}
\setlength{\intextsep}{5pt plus 2pt minus 2pt}

\title{Vortex pinning and flux flow microwave studies of coated conductors}

\author{N.~Pompeo,~\IEEEmembership{Senior Member,~IEEE,}
        K. Torokhtii,~\IEEEmembership{Member,~IEEE,} 
        A. Alimenti,~\IEEEmembership{Student Member,~IEEE,} 
        A. Mancini,
        G. Celentano,
        E.~Silva,~\IEEEmembership{Senior Member,~IEEE}%
\thanks{A. Alimenti, N. Pompeo, E. Silva and K. Torokhtii are with the Department
of Engineering, Universit\`{a} Roma Tre, 00146 Roma,
Italy. Corresponding author: E. Silva; e-mail: enrico.silva@uniroma3.it.}%
\thanks{A. Mancini and G. Celentano are with the Superconductivity Laboratory, Italian National Agency for New Technologies Energy and Sustainable Economic Development (ENEA), Frascati, Italy.}%
\thanks{Manuscript received October 30, 2018.}}

{}

\maketitle
\copyrightnotice

\begin{abstract}
Demanding microwave applications in a magnetic field require the material optimization not only in zero-field but, more important, in the in-field flux motion dominated  regime. However, the effect of artificial pinning centers (APC) remains unclear at high frequency. Moreover, in coated conductors the evaluation of the  high frequency material properties is difficult due to the complicated electromagnetic problem of a thin superconducting film on a buffered metal substrate.
In this paper we present an experimental study at 48 GHz of \mbox{150--200}~nm YBa$_2$Cu$_3$O$_{7-x}$ coated conductors, with and without APCs, on buffered Ni-5at\%W tapes. By properly addressing the electromagnetic problem of the extraction of the superconductor parameters from the measured overall surface impedance $Z$, we are able to extract and to comment on the London penetration depth, the flux flow resistivity and the pinning constant, highlighting the effect of artificial pinning centers in these samples.

\end{abstract}


\IEEEpeerreviewmaketitle

\section{Introduction}
\label{intro}

\IEEEPARstart{H}{igh} frequency applications of high-$T_c$ superconductors, and consequently the corresponding oriented studies, have been scarce outside of the superconducting electronic field. This reality is recently changing thanks to the maturity that technological materials are achieving, together with the even more demanding requirements for devices like accelerating cavities \cite{cavities}. A similar effect in terms of revamped interest is coming from emerging fields like the hunt for galactic axions \cite{digioacchino}, as well as future implementation of particles colliders \cite{FCC}. 
The dominant approach for making one of the most promising high-$T_c$ superconductors, YBa$_2$Cu$_3$O$_{7-\delta}$ (YBCO), available on large (namely long, in view of cable applications) areas is the so-called coated conductor technology \cite{Senatore2014}. The latter consists in the growing of thin superconductor films on flexible metallic tapes, which provide the mechanical support. Buffers comprising several heterogeneous thin layers are necessary for chemical and physical matching of the materials. Moreover, large efforts have been spent to improve the current carrying capabilities of such superconductors with a technological production compatible approach. Artificial pinning centers (APCs), needed to improve the in-field flux pinning capabilities, are added during the film growth  \cite{apc}.

In overall, the resulting coated conductor emerges as a complex layered structure. 
Hence the measurement and interpretation of its high frequency electrodynamic response, represented by its surface impedance $Z$, 
requires a careful analysis to isolate the contributions coming from the different layers. 
Indeed, the thin superconducting film is not completely opaque to the impinging high frequency electromagnetic (e.m.) field which, therefore, reaches the underlying substrate. The metallic supporting layer, being a conductor, has a high enough conductivity to be appreciable with respect to the superconductor. The resulting measured surface impedance, which is defined as $Z=H_t/E_t$ \cite{Collin}, where $H_t$ and $E_t$ are the tangential components of the e.m. fields on the film surface, arises as the sum of the e.m. wave coming from multiple reflections at the interfaces between layers.

In previous works, we have addressed the problem of the extraction of the superconductor intrinsic impedance and complex resistivity $\tilde{\rho}_s$ from the measured $Z$ on the coated conductor structure in measurements performed both at fixed temperature $T$ and varying static magnetic field $B$ \cite{Pompeo2018} and at zero field and varying temperature \cite{PompeoIMEKO}, demonstrating the feasibility of the developed method. 
In the present manuscript, we intend to exploit the proposed analysis for the study and comparison of the pinning performances of  YBCO based coated conductors, with and without APCs. We will focus on the main vortex parameters and on their {$T$-}dependence.

The paper is organised as follows: in Section \ref{sec:experiment} we first describe the samples and the experimental setup, then we will briefly recall the high frequency electrodynamic model which includes the vortex motion parameters of interest. In Section \ref{sec:results}, the measurements will be presented and analysed. Finally the conclusions will be drawn in Section \ref{sec:conc}.

\section{Experimental setup and method}
\label{sec:experiment}
Two YBCO thin films, of nominal thickness $t_s=175\;$nm, with and without the addition of 5\% BaZrO$_3$ (BZO), have been produced by Pulse Laser Deposition on Ni-5at.\%W tapes provided by EVICO. 
A stack of buffer layers comprising CeO$_2$/YSZ/CeO$_2$/Pd with thicknesses (15/110/45/200)~nm, respectively, has been deposited to provide chemical and mechanical stability and to improve the epitaxial growth of the superconducting film \cite{Mancini2008}. 
The two films are similar with respect to the structural properties as resulting from X-ray diffraction analysis reported in \cite{Torokhtii2016}. More details about film deposition process can be found in \cite{Mancini2008} and \cite{Silva2015}. The 
multilayer structure of the samples is sketched in the inset of Fig. \ref{fig:trans}.

The surface impedance $Z$ measurements have been performed by means of a dielectric resonator used in the end-wall replacement configuration. 
By exciting the TE$_{011}$ mode (resonant frequency $f_{res}\sim47.9\;$GHz) through a two-port connection (transmission operation), {the frequency dependent resonant curve is measured and fitted, yielding}
the unloaded quality factor $Q$ and $f_{res}$ \cite{PompeoMSR2014} from which, through standard e.m. theory \cite{Staelin}, $Z$ is computed as: 
\begin{eqnarray}
\label{eq:DZmeas}
    \Delta Z=\Delta R+\rmi\Delta X=G_s\left(\Delta\frac{1}{Q}-2\rmi\frac{\Delta f_{res}}{f_{res,ref}}\right)
\end{eqnarray}
\noindent where $\Delta A=A(x)-A(x_{ref})$ represents the variation of the quantity $A$ with respect to the external parameter $x$, which stands for either the temperature $T$ or the static magnetic field $H$, and 
$G_s$ is a calculated geometrical factor.
The measurement system comprises a liquid/solid nitrogen cryostat, which allows to vary the temperature in the range $60\;$K--$T_c$, and a conventional electromagnet which provides magnetic fields up to $\sim0.8\;$T applied perpendicularly to the sample surface, hence parallel to the YBCO film $c$-axis. We remark that the present system is adequate for measurements of the field--variation of $Z$, while high--sensitivity measurements near $R=0$, e.g. of the residual surface resistance, are out of its capabilities.

To interpret the surface impedance measurements, {the above described layered structure of the sample must be taken into account.}
Resorting to the transmission line formalism \cite{Collin}, 
the overall effective $Z_{eff}$ is computed with the recursive formula:
\begin{equation}
\label{eq:Zeff} 
Z^{(i)}_{eff}=Z^{(i)}_{char}\frac{Z^{(i-1)}_{eff}+\rmi Z^{(i)}_{char}\tan(k^{(i)} t^{(i)})}{Z^{(i)}_{char}+\rmi Z^{(i-1)}_{eff}\tan(k^{(i)} t^{(i)})}
\end{equation}
where $Z^{(i)}_{char}$, $k^{(i)}={2\pi f\mu_0}/{Z^{(i)}_{char}}$ and $t^{(i)}$ are, respectively, the characteristic surface impedance, the wave propagation constant and the thickness for the $i^{th}$ layer; $Z^{(i-1)}_{eff}$ is the overall effective surface impedance of the underlying layer stack, having the layer $(i-1)^{th}$ on top; $f$ is the frequency and $\mu_0$ is the vacuum magnetic permeability.
The characteristic impedance $Z_{char}$ is equal to the surface impedance expression $Z_{bulk}$ for bulk {materials,} $\sqrt{\rmi2\pi f\mu_0\rhos}$ and $\sqrt{\mu_0/(\varepsilon_0\varepsilon_r)}$ for a (super)conductor with complex resistivity $\rhos$, in the local limit, and for an insulator with relative permittivity $\varepsilon_r$, respectively. 
It is useful to recall here that when a superconducting thin film is grown on a thick, insulating substrate, as opposed to the buffered metal here considered, the effective surface impedance simplifies {to the} so called thin-film expression $Z_{film}=\rhos/t_s$ \cite{thinfilm} at the condition, usually verified, that $t_s\ll min(\lambda, \delta_n)$, with $\lambda$ and $\delta_n$ the {London and} the normal fluid penetration depth, respectively.
Considering the coated conductor structure and the condition $t_s\ll min(\lambda, \delta_n)$, Eq. \eqref{eq:Zeff} can be cast in a more readable form:  
\begin{equation}
\label{eq:Zeff_approx} 
Z=Z_{film} \frac{\Zsub+\rmi 2\pi f\mu_0 t_s}{Z_{film} + \rmi \Zsub}
\end{equation}
which highlights that, apart from the often negligible term $\rmi 2\pi f\mu_0 t_s$, the coated conductor effective surface impedance $Z$ comes out as the parallel between the superconductor film $Z_{film}$ and the substrate $\Zsub$ impedances.

The superconductor $\rhos$ can be written as \cite{CC}:
\begin{equation}
\label{eq:rhoCC}
     \rhos=(\rho_{vm}+\rmi/\sigma_2)/(1+\rmi\sigma_1/\sigma_2)
\end{equation}
which includes both the two-fluid conductivity $\sigma_f=\sigma_1-\rmi \sigma_2$ and the vortex motion resistivity $\rho_{vm}$.
Textbook models \cite{Tinkham} allow to write $\sigma_f={x_n(t)}/{\rho_n}-\rmi/({\omega \mu_0\lambda^2(T))}$ with $\lambda(T)={\lambda_0}/\sqrt{x_s(T)}$, being $\lambda_0$ the zero temperature $\lambda$, $\rho_n$ the normal state resistivity and $x_n(T)+x_s(T)=1$ the normalized normal and super- fluid fractions. Typical temperature dependence for high-$T_c$ superconductors can be taken as $x_s(T)=1-(T/T_c)^2$.

{In order to discuss the vortex motion response, we must choose a proper model. In the microwave frequency range, the first important distinction is between two alternative dynamic regimes. In the low frequency region, the disorder dominated vortex dynamics is characterized by various energy/time scales which manifest itself with various vortex creep rates, leading to a response typical of the vortex glass, in which the vortex motion resistivity has a power law dependence on the frequency \cite{belk, wu}. In the higher frequency range, on the other hand, motion of vortices in the bottom of the pinning wells is so fast and short-ranged that multi-scale creep is no longer dominant and the dynamics is well described through the combined effect of viscous drag and a single characteristic pinning frequency \cite{wu, tsuchiya2001}. The crossover frequency separating the two regimes is, in the ``high'' temperature range here considered, smaller than 10 GHz \cite{wu}, well below our measuring frequency, ~48 GHz. Hence,} neglecting thermal creep effects, relevant only near $T_c$, without loss of generality \cite{PompeoPRB08} we can resort to the basic, single-characteristic frequency proposed by Gittleman and Rosenblum \cite{GR}: 
\begin{equation}
\label{eq:rhoGR}
    \rho_{vm}=\rho_{vm1}+\rmi\rho_{vm2}=\frac{\Phi_0B}{\eta}\frac{1}{1-\rmi\frac{f_p}{f}}
\end{equation}
where $\Phi_{0}$ is the flux quantum, $\eta$ is the vortex viscosity (connected to the quasiparticle properties around and near the vortex cores and related to the flux flow resistivity by $\rho_{ff}={\Phi_0B}/{\eta}$), and $f_p={k_p}/{2\pi\eta}$ is the depinning frequency, marking the separation between the pinning-dominated, low frequency Campbell regime and the dissipation-dominated, high frequency flux flow regime. 
The pinning constant $k_p$, also known as Labusch parameter, is a measure of the pinning force strength acting on vortices when they are ``infinitesimally'' displaced from their equilibrium positions. We stress that this approximation is valid at microwave frequencies, where  the vortex oscillation amplitudes are well below 1 nm.

We note that, within this model, the adimensional ratio $r=\rho_{vm2}/\rho_{vm1}=f_p/f$ represents the normalized depinning frequency. This important parameter, when measuring thin films over insulating substrates, is obtained directly as \mbox{$r=\Delta X(H)/\Delta R(H)$}.
On the other hand, when measuring coated conductors, the normalized depinning frequency cannot be derived so straightforwardly \cite{Pompeo2018}, but requires a careful analysis of the data with the above recalled model. 

\section{Results and Discussion}
\label{sec:results}
\begin{figure}[hbt!]
\centering
\centerline{\includegraphics[scale=\myscalefactor]{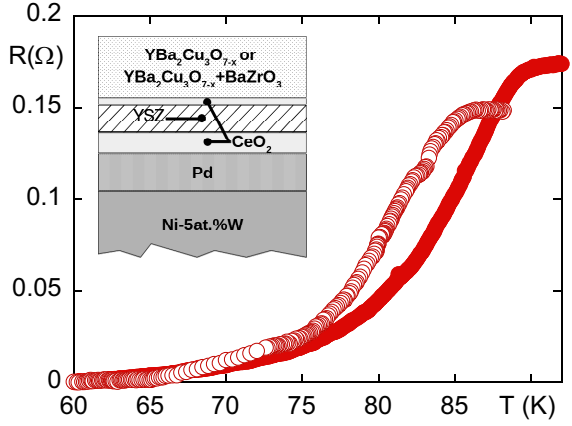}}
\caption{
Measured $R$ vs $T$ for the YBCO (full dots) and YBCO/BZO (open circles) samples. Inset: multilayer structure of the coated conductor.
}
\label{fig:trans}
\end{figure}

We first show the zero field resistive transition at 48 GHz measured on both samples,
Fig. \ref{fig:trans}. 
We take $R(T_{min})\sim0$ as reference value, which does not introduce significant errors, as discussed in Sec. \ref{sec:experiment}. Thus in the temperature range here considered one has $\Delta R(T)\simeq R(T)$ apart from very near to $T_{min}=60\;$K. 
As it can be seen, the addition of BZO reduces the critical temperature to 86 K, with respect to the 90 K exhibited by the pure YBCO sample, an effect already reported in literature \cite{Cantoni, Huhtinen}.
The normal state values are $R_n=R(T>T_c)=0.174\;\Omega$ and $R_n=0.148\;\Omega$, for YBCO and YBCO/BZO, respectively. 
This difference in the normal state values cannot be directly attributed to the superconducting films since, as extensively discussed in our previous works  \cite{Pompeo2018, PompeoIMEKO}, the surface resistance of the coated conductor above $T_c$ is dominated by the effective substrate $\Rsub$, i.e. $R_n\simeq\Rsub$.

\begin{figure}[hbt!]
\centering
\centerline{\includegraphics[scale=\myscalefactor]{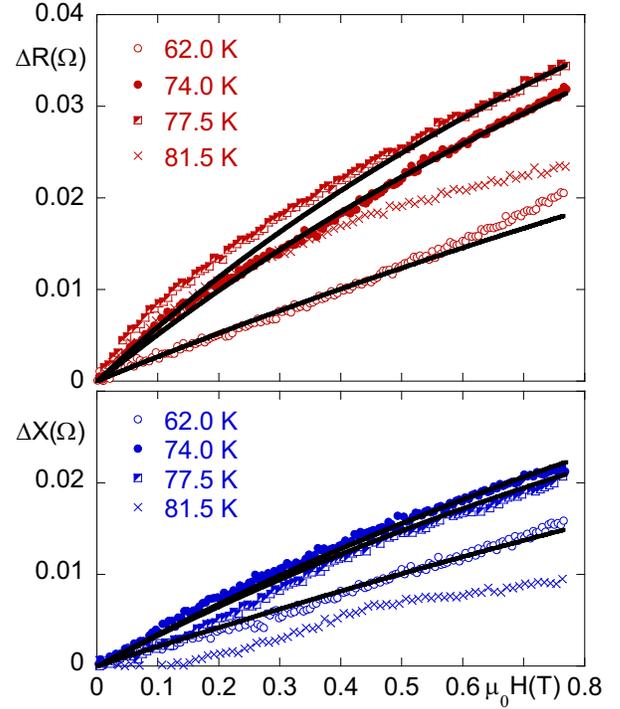}}
\caption{
Field variation of the $\Delta R(H)$ and  $\Delta X(H)$, upper and lower panel, respectively, at selected temperatures for sample YBCO/BZO.}
\label{fig:sweep}
\end{figure}

\begin{figure}[hbt!]
\centering
\centerline{\includegraphics[scale=\myscalefactor]{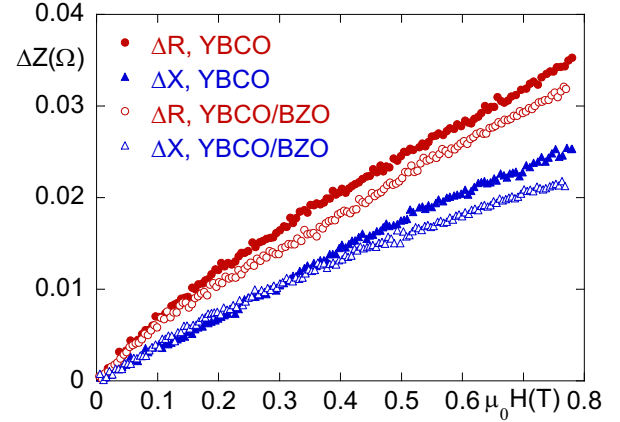}}
\caption{
Field variation of $\Delta Z(H)$ at the same reduced temperature $t=T/T_c\simeq0.86$ for sample YBCO (open symbols) and YBCO/BZO (full symbols). Circle and triangles are for $\Delta R(H)$ and $\Delta X(H)$, respectively.}
\label{fig:reduced}
\end{figure}

We now focus on the surface impedance measurements performed at fixed temperature by varying the field. 
Field sweeps at selected temperatures are reported in Figs. \ref{fig:sweep} and \ref{fig:reduced}.
It can be seen that $\Delta Z(H)$ is always increasing with the field, with a predominant downward curvature. This sublinear (in $B$) behavior is different from the expectation for of the simple $\rho_{vm}/t_s\propto B$ dependence which would apply in the thin film approximation with field-independent vortex parameters and the vortex resistivity proportional to the number of fluxons. 

The temperature evolution of $\Delta Z(H)$ shows the usual behavior, common to both samples: an increase of the amplitude, followed by a decrease approaching $T_c$, see  Fig. \ref{fig:sweep} for the YBCO/BZO sample. Another important remark is that  $\Delta R>\Delta X$ for both samples at all $T$. In a conventional analysis, this finding would be an indication of weak pinning.
It should be stressed that, once the effect of the different $T_c$ is ruled out \cite{Torokhtii2016}, the pure and nanostructured sample behave analogously: in Fig. \ref{fig:reduced} we plot $\Delta Z(H)$ for both samples taken at the same reduced temperature $t=T/T_c\simeq 0.86$ ($T$=77.5 K and 74 K in YBCO and YBCO/BZO, respectively). It is seen that the data are well within 10\% between different samples.

In order to extract and comment on the main vortex parameters, we now proceed to analyse these raw measurements. As previously discussed in \cite{Pompeo2018}, the determination of $\rhos$ and hence of $\rho_{vm}$ (Eqs. \eqref{eq:rhoCC} and \eqref{eq:rhoGR}) requires the knowledge of the substrate impedance $\Zsub$. 
Since no separate measurements of $\Zsub$ are available, $\Zsub$ must be indirectly determined. Indeed, the substrate multilayer structure can be fully characterized with the (e.m.) model (Eq. \eqref{eq:Zeff}) by using the known nominal thicknesses of the layers, and by taking literature values for their relative permittivities and electrical conductivities. 
In particular, we take $\rho_{Pd}=5\cdot10^{-8}\;\Omega\cdot$m \cite{Shivaprasad1980, Matula1980}, $\rho_{NiW}=25\cdot10^{-8}\;\Omega\cdot$m \cite{Muth1983, Lakshmi2010}, and $\epsilon_r=23$ and $\epsilon_r=27$ for CeO$_2$ and YSZ, respectively \cite{Santha2004, Lanagan1989}. 
Then, by exploiting the model represented by Eq. \eqref{eq:Zeff}, 
we can reproduce the value of $R(T>T_c)=R_n\simeq\Rsub$ with very small changes to the literature values for the Pd parameters. 

In fact, the computed value for $\Rsub$ is mostly sensitive to the Pd layer properties, $\rho_{Pd}$ and $t_{Pd}$, through their ratio. We chose to adjust only $\rho_{Pd}$ to $6.41\cdot10^{-8}\;\Omega\cdot$m and \mbox{$4.63\cdot10^{-8}\;\Omega\cdot$m}, for YBCO and YBCO/BZO respectively. Identical results would have been obtained by changing the Pd layer nominal thickness by 10--20\%: both results are physically plausible, being within the deposition tolerances, and their exact nature is irrelevant for the rest of the computations. We note that the model recover the correct values for $R_n$ on two samples
by resorting to the adjustment of one parameter only, a point in favour to the validity of the model itself.
Once $\Zsub$ is computed, the measured $\Delta Z(H)$ can be fitted through Eq. \eqref{eq:Zeff_approx} by inserting $\rhos$ in the superconductor surface impedance: the main fit parameters are $\eta$, $k_p$ and $\lambda$, whereas the fit is not sensitive to $\rho_n$ within ample ranges. Hence, the latter is taken equal to the typical value $\rho_n=1\cdot 10^{-6}\;\Omega\cdot$m.

Fit results for YBCO/BZO (similar results, here not shown, are obtained for YBCO) are reported as thick black lines in Fig. \ref{fig:sweep}. It can be seen that field independent parameters (including $\lambda$, which is equivalent to having neglected field pair-breaking effects, acceptable not too close to $T_c$) 
yield good fits, and in particular the model is able to reproduce the downward curvature of the data. Approaching $T_c$, the downward curvature of the measured $\Delta Z$ becomes even more pronounced, and not fully reproducible with the model. The fit could be reconciled with the data by inserting a field--dependent pinning constant, which is a very reasonable feature very close to $T_c$. However, we found that it was not possible to ascertain the exact field dependence, and then we believe that the fit becomes unreliable.
The temperature dependent London penetration depth $\lambda(T)$ is consistent with $\lambda^{2}=\lambda^2_0/(1-t^2)$,
a typical dependence for high-$T_c$ superconductors, with $\lambda_0$ equal to 130 nm and 120 nm, for YBCO and YBCO/BZO, respectively.
We have checked whether magnetic field pair-breaking effects could improve the fit and determined that, in the $T$ and $B$ range under investigation, they are inessential.

\begin{figure}[hbt!]
\centering
\centerline{\includegraphics[scale=\myscalefactor]{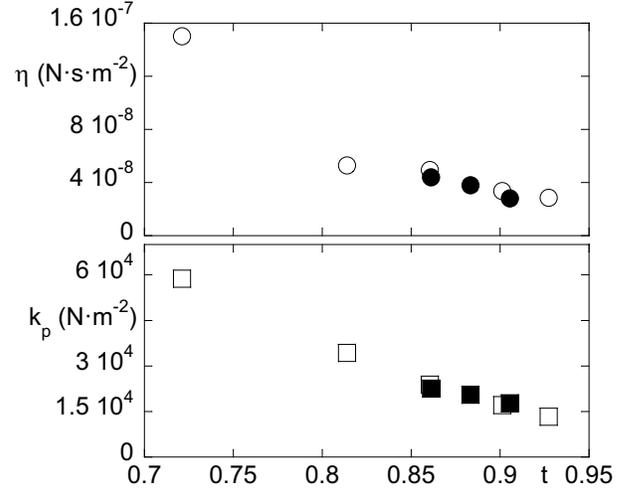}}
\caption{
Fit vortex parameters vs $T$ for YBCO (full symbols) and YBCO/BZO (open symbols). Upper panel: vortex viscosity; lower panel: pinning constant.}
\label{fig:fit_parameters}
\end{figure}

Finally, the fit vortex parameters $\eta(t)$ and $k_p(t)$ are reported in Fig. \ref{fig:fit_parameters}.
Quantitatively, they are consistent with published values reported in literature
obtained in bulk and epitaxial films on insulating substrates \cite{Golosovsky, pompeoAPL2007, pompeoPC2011, pompeoAPL2013, torokhtiiIEEE2017} (for example, at 77 K $k_p\sim5\textendash30\;$kN$\cdot$m$^{-2}$ and $\eta\sim0.5\textendash1.5\cdot10^{-7}$N$\cdot$s$\cdot$m$^{-2}$. To the best of our knowledge, this is the first determination of the vortex parameters with their temperature dependence on coated conductors, a part the single $T=77\;$K data point reported in our previous work \cite{Pompeo2018}.
The slight reduction of $\eta$ with respect to the data on films on crystalline substrate could be an indication, entirely reasonable, for an enhanced scattering of the quasiparticles, perhaps connected to the textured nature of the YBCO films. Since $\eta$ can be related to the normal state resistivity through the Bardeen-Stephen model \cite{BS}, it would be interesting to separately determine $\rho_n$ to further investigate this point. Unfortunately, the very nature of the coated conductor prevents a direct measure of $\rho_n$ at microwaves, so that a separate measurement should be performed. 
On the other hand, $k_p$ is slightly above the average values in {literature} for both samples, so that perhaps the texturing pattern, with its grain boundaries, could play a role here.
As far as the comparison between the two samples is concerned, they present similar values on both $\eta$ and $k_p$, suggesting that the BZO addition, despite the $T_c$ reduction, does not improve appreciably the pinning properties at high frequency {of these particular samples, in stark contrast with what observed in other samples \cite{frolova2016}}. 
{This can be compared with the observed slight increase of the critical current $J_c$ \cite{Torokhtii2016} in the BZO-added sample. These results can be interpreted considering that $J_c$ measures the pinning strength for large displacements of the vortices, up to the detaching distance from the pinning centers. Hence, $J_c$ is a measure of the pinning barriers height. On the other hand, $k_p$ measures their steepness near the bottom. Together, the two characterizations provide complementary information about the pinning effects.
}

Finally, the pinning frequency is $\sim80\;$GHz at t=0.86 for both samples: a quite high figure, which nevertheless arises both due to moderate $k_p$ and to slightly below typical values of $\eta$. A similar phenomenon, although much enhanced, is observed in thin Tl$_2$Ba$_2$CaCu$_2$O$_{8+x}$ films \cite{tallioTAS}.

\section{Summary}
\label{sec:conc}
We have presented extended measurements of the microwave surface impedance of 
YBa$_2$Cu$_3$O$_{7-\delta}$ and YBa$_2$Cu$_3$O$_{7-\delta}$/BaZrO$_3$ coated conductors.
We have highlighted how the complex, multilayer structure of a coated conductor conspire to obfuscate the role of the superconducting material properties in the measurements of the effective surface impedance of the overall structure.
By exploiting a previously developed approach, we extracted
the main vortex parameters, namely the vortex viscosity, the pinning constant and their ratio, the depinning frequency. This result is, to the best of our knowledge, the first determination of these quantities in coated conductors that take into account the coated conductor structure.
The obtained parameters exhibit a conventional temperature dependence and are quantitatively consistent with values known from literature. From a technological point of view, the similar values of $\eta$ and, more importantly $k_p$, between the two samples allow to state that, in the present case, the addition of BaZrO$_3$ has no appreciable effects on the high frequency pinning efficiency.

\section*{Acknowledgment}
This work has been partially supported by EURATOM under an EUROFUSION Enabling Research Grant. The views and opinions expressed herein do not necessarily reflect those of the European Commission.

\ifCLASSOPTIONcaptionsoff
  \newpage
\fi

\end{document}